\documentclass[12pt]{iopart}
\begin{document}
\title[Asymptotic Flatness in Rainbow Gravity]{Asymptotic Flatness in Rainbow Gravity}
\author{Jonathan Hackett}
\address{
Perimeter Institute for Theoretical Physics,\\
31 Caroline St. N., Waterloo, Ontario N2L 2Y5 , Canada, and \\
Department of Physics, University of Waterloo,\\
Waterloo, Ontario N2L 3G1, Canada\\}

\begin{abstract}

A construction of conformal infinity in null and spatial directions
is constructed for the Rainbow-flat space-time corresponding to
doubly special relativity.  From this construction a definition of
asymptotic DSRness is put forward which is compatible with the
correspondence principle of Rainbow Gravity.  Furthermore a result
equating asymptotically flat space-times with asymptotically DSR
spacetimes is presented.

\end{abstract}
\maketitle
\section{Introduction}
The idea of a fundamental length scale has emerged in multiple approaches to quantum gravity.  This length typically identified as the Planck length $l_p$, is expected to be the scale at which quantum gravitational corrections to our present theories would be required.
However, the idea that when probing below the Planck length we will require new physics to describe the resultant phenomena is in direct contradiction with special relativity.  How can we reconcile the idea of a fundamental length scale when special relativity allows lengths to contract?
This apparent paradox of quantum gravity is what gave the impetus behind the original work in doubly special relativity\cite{kowalskidsr, gac3, gac2, GAC1}.  Doubly special relativity fixes the planck length and attempts to occupy the position of a flat space-time limit of quantum gravity.

Recent work on Doubly special relativity has been spurred on by current experiments that could provide a fertile testing ground for its results.  Experiments such as GLAST and AUGER ~\cite{glast, auger} provide the opportunity to test the GZK cutoff and the constancy of the speed of light, both of which are
subjects which DSR is capable of making predictions for.  The other reason for the increased excitement in DSR is the possibility of it providing increased insight into quantum gravity.  The ingredients that lead to DSR are not dependent upon any particular attempt towards a quantum theory of gravity
and in fact are based solely upon attempting to combine the ideas of special relativity and a fundamental length scale.  Due to the simplicity of its construction, it is possible that DSR could provide not only hints into the structure of space-time in a complete theory of quantum gravity, but that it could place restrictions upon one as well.

The picture of space-time resulting from DSR is in some ways still
an open question.  Some approaches to DSR (particularly those that
involve Hopf algebras) have resulted in a picture of space-time
which is non-commutative, these approaches though interesting have
yet to present any physical predictions, primarily due to the
difficulty in construction large-scale behaviour from
non-commutative spacetimes. Fortunately there is an alternative to
non-commutative geometry as the arena in which the space-time of DSR
is understood, by attempting to extend DSR to general relativity
Magueijo and Smolin put forward Rainbow Gravity.  In Rainbow
gravity, the need for non-commutative geometry is avoided by
introducing a metric which 'runs' with respect to the energy at
which it is probed.  Not only does this approach avoid the need for
non-commutative geometry to describe the space-time of DSR, it
allows DSR to be extended to a theory of gravity much like general
relativity.

The intent of this paper is to explore the implications of Rainbow
Gravity further.  Of particular interest is the asymptotic behaviour
of the theory as this is an area where other approaches to DSR have
not yet produced results.  To this end we will investigate the ideas
of conformal infinity and asymptotic DSRness (in analogy with
asymptotically flat space-times in General relativity) in Rainbow
gravity.
\section{Rainbow Gravity}
\label{rainbow}
 Rainbow Gravity \cite{lj} is an attempt at
construction an extension of DSR into a general relativity framework
which has at its foundation the proposal that the geometry of a
space-time 'runs' with the energy scale at which the geometry is
being probed.  The implication of this is that the metric of a
space-time becomes energy dependant.  Fortunately a natural proposal
for the manner in which the metric could vary with respect to energy
emerged from previous work on DSR \cite{leeandjoao1}, which is the
form that we shall use.

Rainbow Gravity is governed by two principles, the correspondence
principle and the modified equivalence principle.   These two
principles are stated as follows ~\cite{lj}

\begin{quote}
\textbf{\textit{Correspondence principle}} \\
In the limit of low energies relative to the Planck Energy, standard
general relativity is recovered.  That is for any rainbow metric
$g_{ab}(E)$ corresponding to a standard metric $g_{ab}$ the
following limit holds true:

\begin{equation}
\lim_{\frac{E}{E_{pl}}\rightarrow 0} g_{ab}(E) = g_{ab}.
\end{equation}
\end{quote}
\begin{quote}
\textbf{\textit{Modified Equivalance principle}}\\
Given a region of space-time with a radius of curvature $R$ such
that
\begin{equation}
R >> E^{-1}_{pl},
\end{equation}
then freely falling observers measure particles and fields with
energies $E$ observe the laws of physics to be the same as modified
special relativity to first order in $\frac{1}{R}$ so long as:
\begin{equation}
\frac{1}{R} << E << E_{pl}.
\end{equation}
Thus they can consider themselves to be inertial observers in a
rainbow flat space-time (to first order in $\frac{1}{R}$) and use a
family of energy dependent orthonormal frames locally given by
\begin{eqnarray}
e_0 = \frac{1}{f(\frac{E}{E_{pl}})}\tilde{e}_0\\ e_i =
\frac{1}{g(\frac{E}{E_{pl}})}\tilde{e}_i,
\end{eqnarray}
with a metric:
\begin{equation}
g(E) = \eta^{ab}e_a\otimes e_b.
\end{equation}
\end{quote}
We shall assume the existence of these two functions $f(E)$ and
$g(E)$ which are strictly greater than zero for small values of $E$;
the small range of restriction is to allow for the possibility that
at significantly greater energies the geometry of space-time could
take on a significantly different character, and the restrictions on
this assumption will be explored further in the context of
asymptotic flatness.  These functions will also be subject to an
additional restriction of invertibility, discussed in section
\ref{citeme2}. It should also be noted that our assumption that
these functions depend only upon the energy, and not momenta is
rooted in the idea that as our family of metrics (instead of just a
single space-time) is to be the dual to the momentum space, the form
of it should not depend on the momenta.

The result of these principles is that the rainbow metric for any govern space-time is actually a family of metrics given by energy-dependent orthonormal frame fields - as presented above -
which must satisfy a `Rainbow Einstein equation'
\begin{equation}
G_{\mu \nu}(E) = 8 \pi G(E) T_{\mu \nu}(E) + g_{\mu \nu} \Lambda(E),
\end{equation}
where Newton's constant and the cosmological constant are now allowed to vary with the energy so long as they obey the correspondence principle.

This is the form of Rainbow Gravity which will be used to study the
idea of asymptotic DSRness in the following sections. It should be
noted that conformal mappings of Rainbow Gravity space-times
pointwise with respect to the energy (at specific energies instead
of treating the energy as a dimension) are possible due to the
similarity between the Rainbow Einstein Equations and the original
Einstein equations.  All Rainbow metrics are actually solutions to
Einstein's equations in a mathematical sense (treating the functions
solely as mathematical concepts, instead of allowing them to
correspond to physical quantities) with the caveat that for energies
where $G(E)$ varies from Newton's constant that the equation is
slightly modified, but not in a manner which would impact the
behaviour of solutions under conformal mappings, nor their
compactifications.

\section{Conformal infinity in Rainbow Minkowski space-times}
\label{infinity} In order to understand the asymptotic behaviour of
space-times in Rainbow Gravity it is useful to have a consistent
manner in which to evaluate quantities 'infinitely far away' in null
and spatial directions.  Additionally in order to be able to ascribe
the title of `asymptotically' DSR to a space-time we need to be able
to identify the behaviour of the flat DSR space-time at these
asymptotic locations.  To do this we shall extend the ideas of $\cal
I$ and $\imath_0$ from asymptotic flatness to DSR, proceeding in a
similar manner to previous demonstrations of these completions of
Minkowski space-time.~\cite{wald}

We begin by considering the components of the metric of the flat DSR
spacetime (we shall call this the deformed Minkowski space-time or
Rainbow space-time) $g_{ab}$
\begin{equation}
ds^2 = \frac{-dt^2}{f^2(E)} + \frac{dr^2}{g^2(E)} +
\frac{r^2}{g^2(E)} d\Omega^2,
\end{equation}
where $d\Omega^2$ is the angular component of the spatial
directions. By setting this equal to zero (and likewise setting the
angular component to zero) we are able to identify the speed of
light as a function of the energy by:
\begin{equation}
c = \frac{dr}{dt} = \frac{g(E)}{f(E)}.
\end{equation}
This allows us to construct (energy dependent) null co-ordinates
given by
\begin{eqnarray}
v = t + \frac{f(E)}{g(E)}r\\
u = t - \frac{f(E)}{g(E)}r,
\end{eqnarray}
and change the metric accordingly to:
\begin{equation}
ds^2 =\frac{1}{f^2(E)}\left( -dvdu + \frac{1}{4} \left(v-u\right)^2
d\Omega^2 \right).
\end{equation}
At this point we see that the only difference between this metric
and the Minkowski space-time metric in null co-ordinates is a factor
of $\frac{1}{f^2(E)}$.  This means that we can perform a conformal
mapping of this space-time into a restriction of the Einstein static
universe by using a conformal factor of
\begin{equation}
\Theta^2 = \frac{4}{f^2(E)\left(1 + v(E)^2\right)\left(1 +
u(E)^2\right)},
\end{equation}
such that the new metric $\hat{g}_{ab}$ is related to the old by:
\begin{equation}
\hat{g}_{ab} = \Theta^2 g_{ab}.
\end{equation}
The only restrictions arising from this being that
\begin{equation}
\label{restriction 1} \frac{f(E)}{g(E)} \neq 0,
\end{equation}
so that our null co-ordinates are well defined for all $E$ and that
\begin{equation}
\label{restriction 2} f(E) \neq 0,
\end{equation}
and that both $\frac{f(E)}{g(E)}$ and $f(E)$ are finite, so that our
conformal factor $\Theta^2$ is well defined for all $E$. This
mapping is made clear by choosing new co-ordinates of:
\begin{eqnarray}
T(E) = \tan^{-1}\left( v(E)\right) + \tan^{-1}\left(u(E)\right)\\
R(E) =\tan^{-1}\left(v(E)\right) - \tan^{-1}\left(u(E)\right).
\end{eqnarray}
The ranges of the new co-ordinates being
\begin{eqnarray}
-\pi < T(E) + R(E) < \pi \\
-\pi < T(E) - R(E) < \pi \\
0 \leq R,
\end{eqnarray}
and the components of the new metric being:
\begin{equation}
\hat{ds}^2 = -dT^2 + dR^2 + \sin^2(R){}d\Omega^2.
\end{equation}
From here we are able to extend the original space-time to the
boundary of the larger space-time to yield an identification of the
`infinity' of the deformed Minkowski space-time as follows:
\begin{quote}
\textbf{Future Null infinity}($\cal{I}^+$) is identified with $T(E) = \pi - R(E)$ for $0 < R < \pi$\\
\textbf{Past Null infinity}($\cal{I}^-$) is identified with $T(E) = -\pi + R(E)$ for $0<R<\pi$\\
\textbf{Spatial infinity}($\imath^0$) is identified with $R(E) = \pi$, $T(E) = 0$.
\end{quote}
We therefore now have an identification of conformal infinity of the
deformed Minkowski space-time with two reasonably physical
restrictions given by equations \ref{restriction 1} and
\ref{restriction 2}.  This allows us to examine asymptotic
properties, and additionally examine the concept of asymptotic
DSRness in curved space-times.

\section{Asymptotically DSR space-times}
\label{citeme2}
 We now wish to build on the identification of
$\cal{I}^+$, $\cal{I}^-$, and $\imath^0$ for deformed Minkowski
space-time by using them to define an `asymptotically DSR
space-time'.  To do this we shall rely on the definitions of
asymptotically flat space-times~\cite{wald, ash1, ash2} and expand
them to incorporate the running metric of Rainbow Gravity.

There is a concern however that as Rainbow Gravity requires two as
of yet unknown functions - $f(E)$ and $g(E)$ - that we cannot assume
that our definitions of $\cal{I}^+$, $\cal{I}^-$, and $\imath^0$
hold for all values of $E$ in the deformed Minkowski space.

We therefore shall define an interval $\chi$ by
\begin{quote}
$\chi$ is the interval $[0, \varphi)$, where $\varphi$ is the
smallest value greater than zero such that when $\frac{E}{E_{pl}} =
\varphi$, $f(E)$ and $g(E)$ fail to satisfy the restrictions given
in section \ref{infinity}.
\end{quote}

We will only address the concept of asymptotic DSRness where
$\frac{E}{E_{pl}}\ \ \epsilon\ \  \chi$ as our concepts of conformal
infinity are ill defined outside of this interval.  This corresponds
to the deterioration of the ability of classical general relativity
to describe the universe in high energy situations corresponding
with short distance phenomena. Such a restriction on the domain in
which we can define asymptotic structures is natural given the
intent of Rainbow Gravity.  It is of course natural to consider
possible values of $\varphi$ in the definition of the interval
$\chi$, and it would be logical to suggest that values of order
$E_{pl}$ are strong candidates for such an upper bound.

The interval $\chi$ also allows us to adequately describe our
restriction of invertibility on $f(E)$ and $g(E)$.  For these two
functions to produce a mapping $U$ from momentum space to itself
satisfying the restrictions outlined in \cite{ljnew},
$\big(Ef(E),pg(E)\big)$ must be invertible on the interval $\chi$,
which therefore adds the additional restriction that
\begin{equation}
g(E) \neq 0 ,
\end{equation}
for all $E>0$.

Within the interval $\chi$, we shall require that for a space-time
to be considered asymptotically DSR at spatial infinity it must
satisfy the following requirements (analogous to those of asymptotic
flatness conditions~\cite{ash1, ash2})

\textbf{Definition 1}
A `rainbow spacetime'
(\textbf{$M$},$g_{ab}(E)$) is considered asymptotically DSR at
spatial infinity if there exists a set of space-times defined by the
parameter $E$ (\textbf{$\hat{M}$}$(E)$,$\hat{g}_{ab}(E)$) where each
space-time is smooth everywhere except at a point $\imath^0(E)$
where $\hat{M}$ is $C^{\>1}$ and $\hat{g}_{ab}$ is $C^{\>0}$, and
that there exists an imbedding of $M(E)$ into its respective
$\hat{M}$ satisfying:
\begin{quote}
\textbf{(\textit{req. 1})} The union of the closures of the causal
future and causal past of $\imath^0(E)$ is equal to the complement
of $M(E)$ in $\hat{M}(E)$, i.e. that:
\begin{equation}
\bar{J^+}(i^0(E)) \cup \bar{J^-}(\imath^0(E)) = \hat{M} - M.
\end{equation}
\textbf{(\textit{req. 2})} There exists a function $\Theta$ on
$\hat{M}$ that is $C^2$ at $\imath^0(E)$ and smooth everywhere else
satisfying
\begin{eqnarray}
\hat{g}_{ab} = \Theta^2 g_{ab}\\
\Theta(\imath^0)=0\\
\hat{\nabla}_a \Theta(\imath^0) = 0\\
\hat{\nabla}_a \hat{\nabla}_b \Theta(\imath^0) =
2\hat{g}_{ab}(\imath^0),
\end{eqnarray}
\end{quote}
for all values of $\frac{E}{E_{pl}} \ \epsilon \ \chi$.

The motivation behind this definition is the desire for rainbow
gravity to be consistent at each value of $E$.  Given any single
value of $E$ within $\chi$ all standard rules of general relativity
should apply and therefore the requirement for a spacetime to be
asymptotically DSR should be that it be able to be mapped to the
deformed Minkowski space in a manner corresponding to the manner in
which asymptotically flat spacetimes are mapped to Minkowski space
through the Einstein static universe.  For a Rainbow spacetime to be
asymptotically DSR however, it must satisfy this requirement at all
energies within the interval $\chi$ however as we desire a
definition which is dependent upon the spacetime, not upon the
specific energy at which it is being probed.  It should be noted
that this requires that only spacetimes which are asymptotically
flat in the low energy limit can be asymptotically DSR.

Expanding this approach, we are able to expand the standard definition of an Asymptotically flat and Empty spacetime as follows:

\textbf{Definition 2} A `Rainbow spacetime'
(\textbf{$M$},$g_{ab}(E)$) is considered asymptotically empty and
DSR at null and spatial infinity if it is asymptotically DSR at
spatial infinity (as defined above) and satisfies
\begin{quote}
\textbf{(\textit{req. 1})} On the union of boundaries of the causal future and causal past of $\imath^0(E)$, $\Theta = 0$ and excepting $\imath^0(E)$, $\hat{\nabla}_a\Theta \neq 0$ on the same.\\
\textbf{(\textit{req. 2})} There exists a neighbourhood $N(E)$ of the union of boundaries of the causal future and causal past of $\imath^0(E)$ in $\hat{M}(E)$ such that $(N(E),\hat{g}_{ab}(E))$ is strongly causal and time orientable, and in the intersection of $N(E)$ and the image of $M(E)$ in $\hat{M}$, $R_{ab}(E) = 0$ (where $R_{ab}(E)$ corresponds to the original physical metric $g_{ab}(E)$)\\
\textbf{(\textit{req. 3})} The map of null directions at $\imath^0(E)$ into the space of inegral curves of $n^a = \hat{g}^{ab}\hat{\nabla}_a\Theta$ on $\cal{I}^+$$(E)$ and $\cal{I}^- $$(E)$ is a diffeomorphism.\\
\textbf{(\textit{req. 4})} For a smooth function, $\omega$, on the complement of $\imath^0(E)$ in $\hat{M}(E)$ with $\omega > 0$ on the union of the image of $M(E)$ in $\hat{M}(E)$ with $\cal{I}^+$$(E)$ and $\cal{I}^-$$(E)$ which satisfies $\hat{\nabla}_a(\omega^4n^a) = 0$ on the union of $\cal{I}^+$$(E)$ and $\cal{I}^-$$(E)$ the vector field $\omega^{-1}n^a$ is complete on  the union of $\cal{I}^+$$(E)$ and $\cal{I}^-$$(E)$.\\
\end{quote}
for all values of $\frac{E}{E_{pl}} \ \epsilon \ \chi$.

These two definitions provide a means of identifying whether those
concepts that are defined through asymptotic behaviour (such as the
energy of an isolated system) carry over to Rainbow Gravity in a
consistent manner.

\section{Correspondence between asymptotic flatness and DSR}

It is a natural question to inquire into whether an asymptotically
flat space-time will - when extended to Rainbow Gravity - correspond
to an asymptotically DSR space-time.  Fortunately we are able to
show that for any metric derived from the conditions outlined in
section \ref{rainbow}, this is the case.

\begin{quote}
\textbf{Theorem 1} Given any Rainbow Gravity metric derived from
orthonormal frame fields
\begin{eqnarray}
e_0 = \frac{1}{f(\frac{E}{E_{pl}})}\tilde{e}_0\\
e_i = \frac{1}{g(\frac{E}{E_{pl}})}\tilde{e}_i,
\end{eqnarray}
with a metric:
\begin{equation}
g(E) = \eta^{ab}e_a\otimes e_b.
\end{equation}
If the metric is asymptotically flat in the limit as
$\frac{E}{E_{pl}}$ goes to zero with a conformal factor of
$\Omega^2$, the Rainbow Gravity metric is asymptotically flat within
the interval $\chi$.
\end{quote}
\begin{quote}
\textbf{Proof} We may freely label the frame fields in such a way
that our metric becomes:
\begin{eqnarray*}
ds^2 = - \frac{A^2(t,x,y,z)}{f^2(\frac{E}{E_{pl}})} dt^2 +
\frac{B^2(t,x,y,z)}{g^2(\frac{E}{E_{pl}})} dx^2+
\frac{C^2(t,x,y,z)}{g^2(\frac{E}{E_{pl}})} dy^2+\\
\frac{D^2(t,x,y,z)}{g^2(\frac{E}{E_{pl}})} dz^2.
\end{eqnarray*}
We then change co-ordinates into a radial form defined by:
\begin{eqnarray*}
x=r cos(\theta)sin(\phi)\\
y=r sin(\theta)sin(\phi)\\
z=r cos(\phi).
\end{eqnarray*}
Our metric in this form (with the dependencies of functions
$A$,$B$,$C$,$D$,$f$,$g$ omitted) becomes:
\begin{eqnarray*}\scriptstyle
ds^2 =- \frac{A^2}{f^2(E)} dt^2 + \frac{1}{g^2} \left[ B^2
cos^2(\theta)sin^2(\theta) + C^2sin^2(\theta)sin^2(\phi) +D^2
cos^2(\phi)\right] dr^2 \\\scriptstyle
+\frac{r^2}{g^2}\left[B^2sin^2(\theta)sin^2(\phi)
+C^2cos^2(\theta)sin^2(\phi)\right]d\theta^2\\\scriptstyle
+\frac{r^2}{g^2}\left[ B^2 cos^2(\theta)cos^2(\phi) + C^2
sin^2(\theta)cos^2(\phi) + D^2 sin^2(\phi)\right]
d\phi^2\\\scriptstyle +  \frac{r}{g^2}\left[-2B^2
cos(\theta)sin(\theta)sin^2(\phi) + 2C^2
sin^2(\phi)sin(\theta)cos(\theta)\right] dr d\theta \\\scriptstyle +
\frac{r}{g^2}\left[ 2B^2
cos^2(\theta)sin(\phi)cos(\phi)+2C^2sin^2(\theta)cos(\phi)sin(\phi)
- 2D^2 sin(\phi)cos(\phi) \right] dr d\phi \\ \scriptstyle+
\frac{r^2}{g^2}\left[-2B^2cos(\theta)sin(\theta)cos(\phi)sin(\phi)
+2C^2cos(\theta)sin(\theta)cos(\phi)sin(\phi)\right]d\theta d\phi.
\end{eqnarray*}
We now find the speed of light by setting the distance to zero
\begin{eqnarray*}
\frac{dr}{dt} = \frac{gA}{fF},
\end{eqnarray*}
where the function $F$ is given by
\begin{eqnarray*}\scriptstyle
F^2 = B^2 cos^2(\theta)sin^2(\theta) + C^2sin^2(\theta)sin^2(\phi)
+D^2 cos^2(\phi),
\end{eqnarray*}
and is by that nature energy independent.  Similarly, we can define
the functions  $H$,$K$,$L$,$M$, and $N$ to be
\begin{eqnarray*}\scriptstyle
H^2 = B^2sin^2(\theta)sin^2(\phi)
+C^2cos^2(\theta)sin^2(\phi)\\\scriptstyle K^2 =  B^2
cos^2(\theta)cos^2(\phi) + C^2 sin^2(\theta)cos^2(\phi) + D^2
sin^2(\phi)\\\scriptstyle L^2 = -2B^2
cos(\theta)sin(\theta)sin^2(\phi) + 2C^2
sin^2(\phi)sin(\theta)cos(\theta)\\\scriptstyle M^2 =  2B^2
cos^2(\theta)sin(\phi)cos(\phi)+2C^2sin^2(\theta)cos(\phi)sin(\phi)
- 2D^2 sin(\phi)cos(\phi) \\\scriptstyle N^2 =
-2B^2cos(\theta)sin(\theta)cos(\phi)sin(\phi)
+2C^2cos(\theta)sin(\theta)cos(\phi)sin(\phi),
\end{eqnarray*}
where each of these functions is also energy independent. This
allows us to write the metric more succinctly as:
\begin{eqnarray*}
ds^2 = - \frac{A^2}{f^2(E)} dt^2 + \frac{F^2}{g^2} dr^2 +
\frac{r^2H^2}{g^2} d\theta^2 + \frac{r^2K^2}{g^2} d\phi^2\\ +
\frac{rL^2}{g^2}drd\theta + \frac{rM^2}{g^2}drd\phi +
\frac{r^2N^2}{g^2}d\theta d\phi.
\end{eqnarray*}
Thus allowing us to much more easily convert to null co-ordinates
(again the energy dependence of these will be omitted in formulas)
\begin{eqnarray*}
u = t - \frac{fF}{gA}r\\
v = t + \frac{fF}{gA}r,
\end{eqnarray*}
and making our metric in the $\left(u,v,\theta,\phi,\right)$
co-ordinates:
\begin{eqnarray*}
ds^2 = - \frac{A^2}{4f^2(E)} (du + dv)^2 + \frac{A^2}{4f^2(E)} (dv -
du)^2 \\+ \frac{A^2H^2}{4F^2f^2} (v-u)^2 d\theta^2 +\frac{A^2K^2}{4F^2f^2} (v-u)^2 d\phi^2\\
+ \frac{A^2L^2}{4F^2f^2} (v-u)(dv - du)d\theta +
\frac{A^2M^2}{4F^2f^2} (v-u)(dv - du)d\phi \\+
\frac{A^2N^2}{4F^2f^2} (v-u)^2d\theta d\phi.
\end{eqnarray*}
However, in this form the energy dependence is entirely within the
co-ordinates and a factor of $\frac{1}{f^2(E)}$ common to all terms.
This means that,within the interval $\chi$, the metric differs by a
conformal factor from the metric of the space in the low energy
limit.  We can therefore find a conformal factor for our energy
dependent space (which we will call $\Omega ^2 (E)$) by making a
conformal mapping of our rainbow space-time to an analogue of the
original asymptotically flat space-time (but with energy dependent
co-ordinates) and then composing this with a mapping by the
conformal factor of $\Omega ^2$.  The resultant conformal factor
obeys all of the requirements for the space to be asymptotically DSR
(as the energy dependence of the co-ordinates has no impact on the
pointwise definition of asymptotic DSRness), and the rainbow
space-time is therefore asymptotically DSR with conformal factor:
\begin{eqnarray*}
\Omega ^2 (E) = \frac{1}{f^2(E)} \Omega ^2.
\end{eqnarray*}
\end{quote}

We therefore have a multitude of examples of asymptotically DSR
space-times.  It furthermore means that due to the conditions of the
definition of asymptotic DSRness given above, the set of
asymptotically DSR space-times is in a 1-1 correspondence with the
set of asymptotically flat space-times.
\section{Conclusion}

The process of conformally mapping DSR into the Einstein static
universe allowed for a natural recreation of the definitions of the
conformal infinities and allowed the definitions for asymptotic DSR
behaviour to follow from natural requirements.  These definitions,
though highly stringent in their point-wise nature are put forward
as minimal (at least given the current knowledge of the functions
$f(E)$ and $g(E)$) should Rainbow Gravity undergo further refinement
it could be possible that these restrictions could be relaxed based
upon the running of the metric being `well behaved'.  These
restrictions correspond directly with the concept of an energy
bound, relating to the origins of Rainbow Gravity from doubly
special relativity and therefore play closely in the defining of
asymptotically flat space-times.

The direct correspondence between asymptotically flat and
asymptotically DSR space-times is an interesting and powerful result
allowing work in general relativity to be applied to Rainbow Gravity
with little work.

Further research into asymptotic structure in Rainbow Gravity,
particularly concerning any modification of the asymptotic
symmetries of the space-time would be a logical next step from this
work, along with investigations of the asymptotic definitions of
energy and angular momentum.

\section{Acknowledgements}
The author is extremely grateful to Lee Smolin for his guidance and
advice throughout this work and to the Perimeter Institute for
providing the position through which this work was carried out.

\end{document}